\documentclass[12pt,a4paper]{iopart}
\pdfoutput=1

\usepackage{graphicx, epsfig,epstopdf}
\usepackage{bm}
\usepackage{mathrsfs}
\usepackage{float}

\newcommand{\ket}[1]{\left\vert#1\right\rangle}

\newcommand{\eqref}[1]{(\ref{#1})}
\newcommand{\ex}[1]{\ensuremath{\langle #1 \rangle}}
 
\begin{document}
\title{Entanglement properties of spin models in triangular lattices}
\author{M. Moreno-Cardoner$^1$}
\address{$^1$ Departament de F\'isica, Universitat Aut\`{o}noma de Barcelona, E08193
Bellaterra, Spain}
\author{S. Paganelli$^2$}
\address{$^2$ International Institute of Physics, Universidade Federal do Rio Grande do Norte, 59012-970 Natal, Brazil}
\author{G. De Chiara$^3$}
\address{$^3$ Centre for Theoretical Atomic, Molecular and Optical Physics,
School of Mathematics and Physics, QueenÕs University, Belfast BT7 1NN, United Kingdom}
\author{A. Sanpera$^{4,1}$}
\address{$^4$ Instituci\'o Catalana de Recerca i Estudis Avan\c{c}ats, E08011
Barcelona}
\address{$^1$ Departament de F\'isica, Universitat Aut\`{o}noma de Barcelona, E08193
Bellaterra, Spain}
\ead{Anna.Sanpera@uab.cat}

\begin{abstract}
The different quantum phases appearing in strongly correlated systems as well as their transitions are closely related to the entanglement shared between their constituents. In 1D systems, it is well established that the entanglement spectrum is linked to the symmetries that  protect the different quantum phases. This relation extends even further at the phase transitions where a direct link associates the entanglement spectrum to the conformal field theory describing the former. For 2D systems much less is known. The lattice geometry becomes a crucial aspect to consider when studying entanglement and phase transitions. Here, we analyze the entanglement properties of triangular spin lattice models by considering also concepts borrowed from quantum information theory such as geometric entanglement.
\end{abstract}


\maketitle

\section{Introduction}
Strongly correlated lattice systems have received considerable attention from the quantum information community due to the intricate nature of their quantum correlations  \cite{Amico08}. Indeed the study of entanglement in these systems has revealed more information than it was anticipated. Within this framework, the analysis of bipartite entanglement, described for example by Renyi entropies or the entanglement entropy (i.e. the Von Neumann entanglement between  blocks), concurrence and negativity, have proven to successfully detect quantum phase transitions. 

To date, most of studies have concentrated on one dimensional (1D) lattices, either in spin or Hubbard models where it was first recognized that at criticality (for continuous quantum phase transitions) the entanglement entropy diverges logarithmically with the size $n_{A}$ of the block obtained in the partition $A/B$ as $\sim c \log n_{A}$, where $c$ is the central charge of the conformal field theory (CFT) describing the critical point \cite{Vidal03}. Out of criticality, the system is no
longer entirely constricted by the central charge but has a dependence on the specific perturbations 
tuned to move from the transition point. It is in this regime, where the
entanglement spectrum $\{\lambda_{i}\}$, i.e., the eigenvalues of the reduced density matrix $\rho_{A}$ obtained after the bipartite cut of the system, provides information that is not included in the entanglement entropy $EE$ being this quantity a single number \cite{Li08, Lepori13}.  

Many-body strongly correlated systems in two dimensional (2D) geometries are comparatively much less studied due to the difficulties in solving non-integrable large systems using e.g. exact diagonalization, quantum Monte Carlo or tensor networks. These systems also incorporate intrinsic novel phenomena, a paradigmatic example being geometric frustration. The study of frustrated 2D lattice systems has received an impulse in the last years from the exciting experimentally advances in the area of quantum simulation with trapped neutral atoms and ions. Nevertheless, entanglement studies have also shed new light on some 2D systems, as for instance in topologically ordered ground states that can be identified by the topological entropy, a subleading term arising in the entanglement entropy that is invariant to the size of the blocs in the partition \cite{Kitaev06,Melko11}.

In this paper we focus on the entanglement properties displayed in triangular spin lattice models, both for spin-1/2 and spin-1 particles.  Our study relies both on exact diagonalization (ED) for relatively small systems up to 12 sites as well as in the results obtained by the recently introduced cluster mean field (CMF) approach, a method we have previously explored to determine the phase diagram of the Heisenberg spin-1 model in the presence of a uniaxial anisotropy \cite{Yamamoto09,Yamamoto3,Moreno-Cardoner14}. To gain a further insight into the different phases appearing in the above models, we analyze the entanglement embedded in their ground states. Entanglement properties, as displayed for instance in the entanglement spectrum, are strongly linked to the underlying symmetries of the model. While this fact has been already recognized in several seminal works in 1D systems \cite{Pollmann10,XGWen11,XGWen13}, in 2D systems the structure displayed by the entanglement spectrum is much more complex and has also a clear dependence on the way the partition is made. Finally, we examine also here the geometric entanglement, i.e. the ``distance'' of the many-body ground state to its closest (unentangled) product state. This quantity, in contrast with the previous ones, contains all possible kinds of entanglement from bipartite to multipartite and behaves as a good entanglement measure. All the above quantities are very sensitive to phase transitions and a proper understanding of their properties is of major importance to gain a deeper understanding of the ways in which matter organizes in nature. 

The structure of our manuscript is as follows: In section II we review the two models under consideration, namely the spin 1/2-XXZ and the spin-1 Heisenberg model in the presence of a uniaxial anisotropy both in a triangular lattice. The triangular lattice is particularly interesting since it leads to geometric frustration in the antiferromagnetic case. We briefly review the phase diagram associated to each model as well as the methods we use to find the ground state of the system. It is worth recalling that that the spin-1 model for large negative values of the uniaxial anisotropy reduces effectively to the spin 1/2-XXZ model. Indeed, this is faithfully reflected in the entanglement properties, which become identically in both cases. In Section III we present and explore the relation between quantum phases and entanglement properties, motivating first our choice of entanglement indicators. Particular emphasis is made here in linking the degeneracies emerging in the entanglement spectrum to the underlying symmetries of the system, whose emergence (vanishing) strongly depends on the way the partition is made. We summarize our results in Section IV.

\section{The triangular lattice: Models and Methods}
\label{Sec:Models}
The triangular lattice is particularly interesting since it leads to geometric frustration in antiferromagnetic spin-$1/2$ models. The spin-$1/2$ XXZ Hamiltonian can be written as follows
\begin{equation}
H_{XXZ}=\sum_{\ex{i,j}}\left[ J_\perp \left(\sigma^x_i \sigma^x_j + \sigma^y_i \sigma^y_j\right) + J_z \sigma^z_i \sigma^z_j\right]
\label{Eq:XXZmodel}
\end{equation}
where $\sigma^{\alpha}$ ($\alpha=x,y,z$) are the usual spin-$1/2$ Pauli matrices and $\ex{i,j}$ refers to all possible pairs of first neighbors. For convenience of comparison with the spin-$1$ model, the parameters in the Hamiltonian are chosen without loss of generality as $J_\perp=(2\cos\theta-\sin\theta)/2$ and $J_z=\sin\theta /2$, the angle  $\theta$ fixing  the ratio between the two terms. 

The Hamiltonian (\ref{Eq:XXZmodel}) is $SO(2)$ symmetric (rotationally invariant around the $\hat{z}$-axis) and it also possesses $Z_2$ time-reversal symmetry. Moreover, it is $SU(2)$ symmetric at the two  Heisenberg points ($\theta=\pi/4, 5\pi/4$). As we will discuss later, these symmetries will be reflected in the entanglement properties of the ground state. 

\begin{figure}[H]
\centering
\includegraphics[width=0.75\textwidth,angle=0]{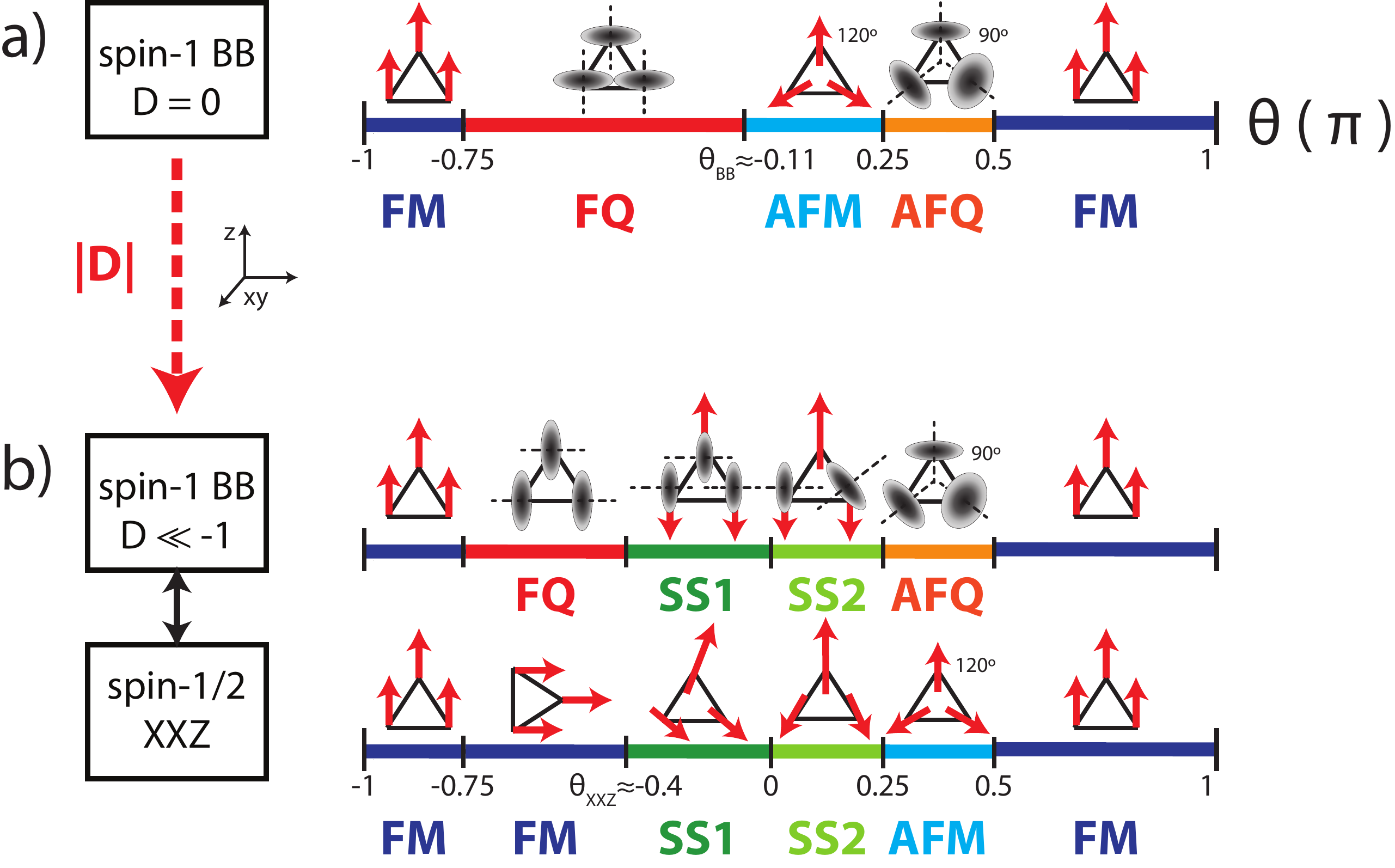}
\caption{Sketch of the phase diagram for the analyzed spin models: a) Spin-$1$ Bilinear-Biquadratic (BB) model at $D=0$, and b) Spin-$1/2$ XXZ model. The spin-$1$ BB model in the very large easy-axis anisotropy limit ($D\ll -1$) exactly maps onto the spin-$1/2$ XXZ model. Red arrows denote dipolar magnetization characterized by 
finite $\ex{\vec{S}}$ and broken lines are the vector directors associated to quadrupolar order when $\ex{\vec{Q}}$ is non-zero (see text).}
\label{Fig:Figure1}
\end{figure}

This system exactly maps into a system of hard-core bosons that tunnel to the nearest neighbors with amplitude $2J_\perp$ and next-neighbor repulsively interact with amplitude $4J_z$, if the operators are transformed according to $S^{-(+)} \rightarrow b^{ (\dagger)}$. This model has been widely studied by several methods (as variational QMC)  \cite{{spin1/20},{spin1/2a},{spin1/2b},{spin1/2c},{spin1/2d},{spin1/2e},{QMC_spin1/2a},{QMC_spin1/2b}} and at least five different phases are expected to arise:  {\it FerroMagnetic} along $\hat{z}$-axis (FMz) for $\theta\in(\pi/2,5\pi/4)$, {\it FerroMagnetic} in $\hat{x}\hat{y}$-plane (FMxy) for $\theta\in (5\pi/4, \Theta_{XXZ})$, {\it Supersolid} (SS1)  for $\theta \in(\Theta_{XXZ}, 0)$, {\it Supersolid} (SS2) for $\theta \in(0, \pi/4)$, and {\it AntiferroMagnetic} ($120^\circ$\textrm{-N\'eel} in the $\hat{x}\hat{y}$-plane) (AFM) for $\theta\in(\pi/4,\pi/2)$ (see sketch of the phase diagram in Figure \ref{Fig:Figure1}). According to QMC calculations \cite{{QMC_spin1/2a},{QMC_spin1/2b}}, the boundary between the FMxy and SS1 phases is around $\Theta_{XXZ}\approx -0.4 \pi$, and this is expected to be a discontinuous phase transtion. In contrast, the transition  between the two Supersolid phases SS1 and SS2 at $\theta=0$ is expected to be continuous in the thermodynamic limit \cite{spin1/2c}.

In the hard-core boson problem the two Supersolid phases are characterized by long range $\sqrt{3} \times \sqrt{3}$ crystal order coexisting with superfluidity, which dubs into the spin language as long-range $\sqrt{3} \times \sqrt{3}$ magnetic order along the $\hat{z}$-axis and net local magnetization in the perpendicular plane. This transverse  magnetization differs in the two phases: for SS1 it takes the same value at each sublattice $(m_\perp,m_\perp,m_\perp)$, whereas for SS2 it corresponds to a configuration $(0,m_\perp,-m_\perp)$.\\

The second model we address here is the spin-$1$ Heisenberg Hamiltonian that reads:
\begin{equation}
H_{BB}(\theta)=\cos\theta \sum_{\ex{i,j}} \vec{S}_i \cdot \vec{S}_j + \sin \theta \sum_{\ex{i,j}} \left(\vec{S}_i \cdot \vec{S}_j\right)^2
\label{Eq:HBBmodel}
\end{equation}
where now ${\vec{S}}=(S_x,S_y,S_z)$ are the usual spin $S=1$ operators and again the sum runs over all possible pairs of first neighboring sites of the triangular lattice. 

This model is the most general (n.n. interaction) rotationally invariant Hamiltonian for spin-$1$, that is, it is invariant when performing the same $SU(2)$ rotation on each of the spins. In general, the two terms in the Hamiltonian compete and can be associated to two different kinds of ordering: dipolar magnetic ordering, emerging when $SU(2)$ symmetry is broken and there is net magnetization, and quadrupolar magnetic order, related to the anisotropy of spin fluctuations. The later also breaks $SU(2)$ symmetry but it preserves $SO(2)$ and $Z_2$ time-reversal symmetries. The two magnetic orderings can be characterized respectively by the two vector order parameters:  $\mbox{\small{\ex{\vec{S}}}}$ and $\mbox{\small{\ex{\vec{Q}}}}=\mbox{\small{\ex{(S_x^2-S_y^2, (2{S}_z^2-S_x^2-S_y^2)/\sqrt{3}, \{S_x,S_y\}, \{S_y, S_z\}, \{S_z,S_x\})}}}$. With this definition, the quadrupolar vector $\vec{Q}$ satisfies ${\small{ (\vec{S}_i \cdot \vec{S}_j)^2= (\vec{Q}_i \cdot \vec{Q}_j - \vec{S}_i \cdot \vec{S}_j)/2+4/3}}$. In a quadrupolar phase, the vector director perpendicular to the plane where the spin maximally fluctuates can also be used to characterize the ordering.

\begin{figure}[H]
\centering
\includegraphics[width=0.60\textwidth,angle=0]{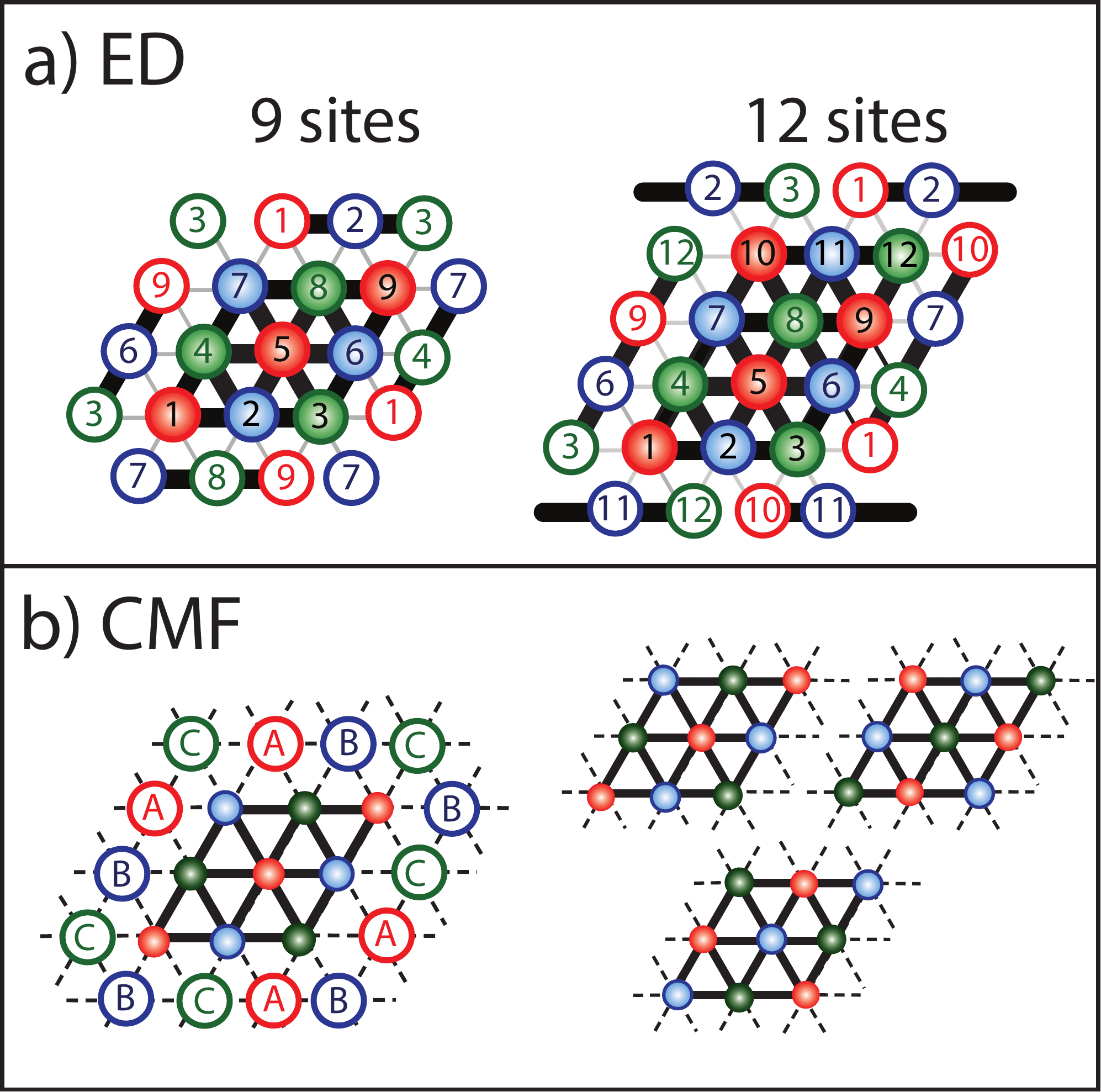}
\caption{ (Color Online). Cluster sizes and geometries used in:  (a) Exact diagonalization with boundary conditions as depicted, (b) Cluster Mean Field approach. Different colors indicate the three-sublattice structure. Solid lines represent quantum mechanical bonds and dashed lines the mean-field interaction between clusters. This acts as an external field (indicated by letters) and it is calculated as an average over all sites that belong to the same sublattice and over all non-equivalent cluster solutions. Right-bottom: non-equivalent cluster configurations for 9 sites.}
\label{Fig:Figure2}
\end{figure}

The phase diagram of the spin-$1$ Heisenberg model in a triangular lattice is well established and has been studied by means of the Gutzwiller ansatz \cite{Toth2011} and Exact Diagonalization of large clusters \cite{Lauchli06}. Here there are four locally ordered phases, which always reproduce the three-sublattice structure inherited from the triangular lattice (see Figure \ref{Fig:Figure2}): {\it FerroMagnetic} (FM) for $\theta\in(\pi/2,5\pi/4)$, {\it FerroQuadrupolar} (FQ) for $\theta\in (5\pi/4, \Theta_{BB})$, {\it AntiferroMagnetic} (AFM) for $\theta \in(\Theta_{BB}, \pi/4)$, and {\it AntiferroQuadrupolar} (AFQ) for $\theta\in(\pi/4,\pi/2)$. Exact diagonalization of large systems \cite{Lauchli06} yield a critical point $\Theta_{BB} \approx -0.11\pi$. In the quadrupolar phases all components of the order parameter $\ex{\vec{S}}$ vanish, while $\ex{\vec{Q}}$ has at least one non-zero component. In the Ferro (F) phases the vectors $\vec{S}$ ($\vec{Q}$) associated to dipolar (quadrupolar) magnetic order are aligned, whereas for the Antiferro (AF) phases they are all contained in the same plane and form $120^\circ$. In the AFQ phase this is equivalent to vector directors on different sublattices that are mutually orthogonal (see sketch of the phase diagram in Figure \ref{Fig:Figure1}).\\

In presence of a uniaxial field, which is quadratic on one spin component, the former model breaks $SU(2)$ symmetry: 
\begin{equation}
H (\theta,D)=H_{BB}(\theta)+D \sum_i (S_i^z)^2
\end{equation}
In the large easy-axis ainsotropy limit ($D\ll -4$), this model directly maps into an effective XXZ spin-$1/2$ model \cite{DeChiara11}, since the the local $\ket{S_i=1, S^{z}_i = 0}$ component is adiabatically suppressed and the two left components per site can be regarded as pseudospin-$1/2$, when mapping $\left\lbrace \ket{1},\ket{-1}\right\rbrace \longleftrightarrow \left\lbrace \ket{\uparrow},\ket{\downarrow} \right\rbrace$. The operators are then related as $\mathcal{P} \vec{S}_i\cdot \vec{S}_j \mathcal{P} = \sigma_i^z \cdot\sigma_j^z$ and $ \mathcal{P} \vec{Q}_i \cdot \vec{Q}_j  \mathcal{P} =2\left(\sigma_i^{x} \sigma_j^{x}+\sigma_i^{y}\sigma_j^{y}\right)+1/3$, and thus, quadrupolar order in the spin-$1$ system corresponds now to transverse magnetization in the spin-$1/2$ system.  

For intermediate values of uniaxial field ($D\approx -4$) the phase diagram is controversial \cite{Bieri2012,Liu2010,Serbyn2011,Xu2012,Moreno-Cardoner14}, and the possibility of a spin-liquid has been suggested using different approaches but not verified. Here we restrict our study to $D=0$ and $D\rightarrow -\infty$ (the spin-$1/2$ XXZ model).\\

Before proceeding further, we summarize here the methods we have used to solve the above models: exact diagonalization (ED)  and the cluster mean field approach (CMF). A detailed survey of the later can be found in \cite{{Yamamoto09},{Yamamoto3},{Moreno-Cardoner14}}. We note that in both methods the geometry of the cluster chosen is very important and determines the symmetries of the ground state. 

Typically we perform ED with clusters of 9 and 12 sites as those displayed in Figure \ref{Fig:Figure2}, and assume the boundary conditions as sketched there. For a triangular lattice the boundary conditions need to be chosen carefully, since they crucially determine the sublattice structure emerging in the ground state. Since we will explore only locally ordered phases that inherit the three-sublattice  structure of the triangular lattice, we choose boundary conditions that allow us to recover this spatial pattern. Thus, for a cluster of 9 sites, it is enough to consider periodic boundary conditions along the primitive vectors of the lattice, whereas for a cluster of 12 sites they will be only periodic along one of the axis (see Figure \ref{Fig:Figure2}).  

In the CMF the lattice is divided into different clusters which in turn are coupled via a mean-field ansatz. This leads to a set of coupled equations that can be self-consistently solved. The way the clusters are chosen (their geometry, number, and how they couple) can strongly influence the result in certain non-trivial phases (as for instance, the Supersolid phases). In fact, there is not a unique way of implementing this approach. Here we impose a three-sublattice structure on the mean-field couplings, which act as an external field over the sites at the cluster boundaries. These external fields are calculated as an average over all sites with the corresponding sublattice order and over all non-equivalent cluster solutions (see Figure \ref{Fig:Figure2}).\\

\begin{figure}[H]
\centering
\includegraphics[width=0.65\textwidth,angle=0]{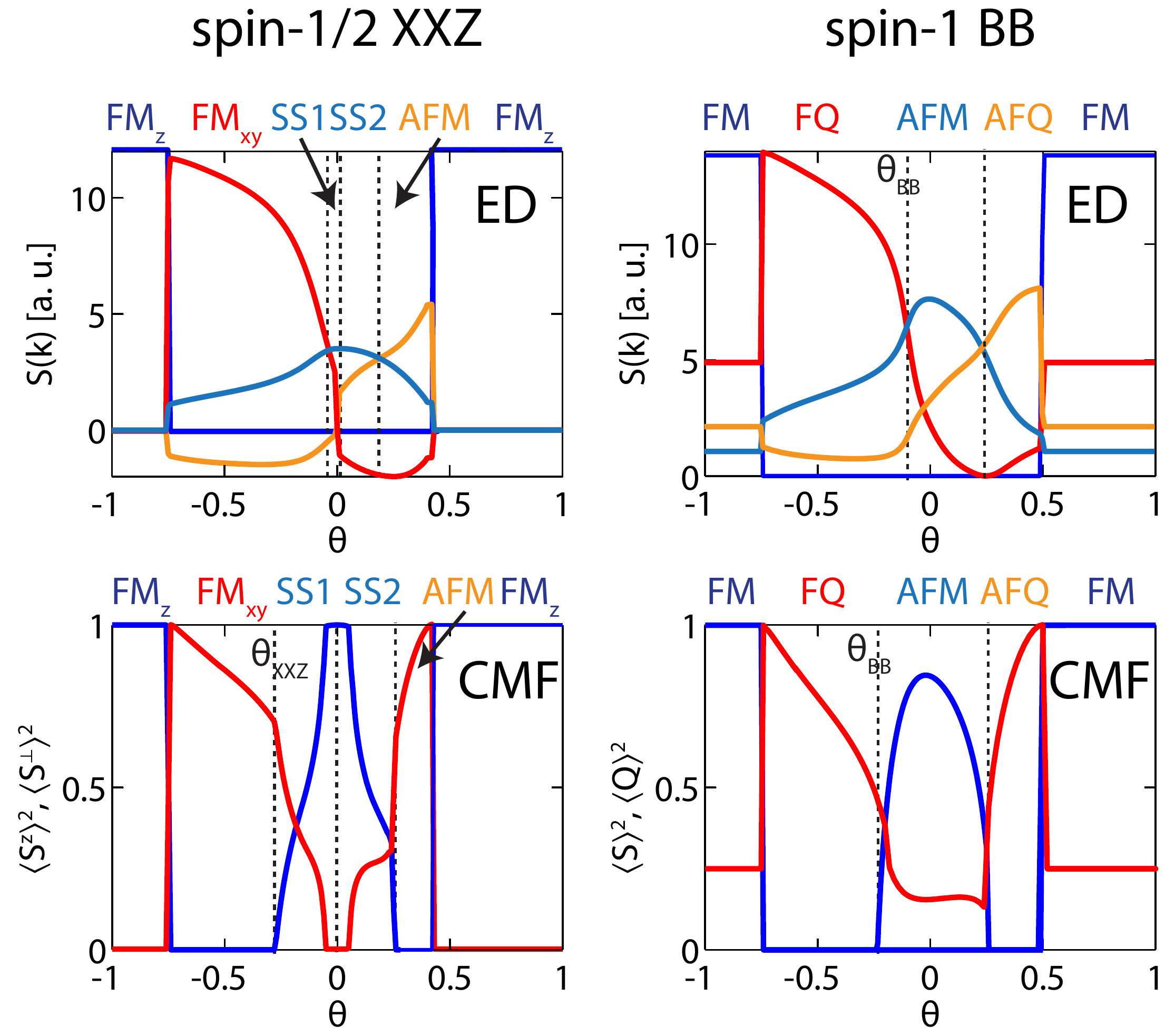}
\caption{(Color Online). Top: Structure factors as a function of  $\theta$ calculated for a 12 sites cluster by exact diagonalization for spin-1/2 (left) and spin-1 (right). Different colors denote different type of ordering  (see text) and phase boundaries are determined when the dominant values cross. Bottom: Local order parameters evaluated in the CMF: $\ex{S^z}^2$ (blue) and $\ex{S^\perp}^2$ (red) in the spin-$1/2$ model (left) and $\ex{\vec{S}}^2$ (blue) and $\ex{\vec{Q}}^2$ (red) in the spin-$1$ model (right).}
\label{Fig:Figure3}
\end{figure}

With these two methods, one obtains the phase diagram of the two spin models shown in Figure \ref{Fig:Figure3}. In the ED solution, in order to determine the phase boundaries, one needs to compare the structure factor corresponding to different types of ordering. This can be defined as:
\begin{equation}
F(\vec{k}) \propto  \sum_j {\exp [i \vec{k}\cdot (\vec{r}_j-\vec{r}_0)] \ex{\vec{A}_0 \cdot \vec{A}_j}}
\end{equation}
where $\vec{A}_i$ is a vector order parameter at site $i$, $\vec{r}_j$ are spatial coordinates of all sites of the cluster and $\vec{r}_0$ a fixed site (any of them). For instance, to compare dipolar and quadrupolar order in the spin-$1$ model one chooses $\vec{A}=\vec{S}$ and $\vec{A}=\vec{Q}$ \cite{Lauchli06}, while phases with magnetization along different direction in the spin-$1/2$ model can be distinguished by choosing  $\vec{A}=S^{\perp}$ and $A=S^{z}$. To evaluate if Ferro or Antiferro- ordering dominates, one needs to compare the corresponding structure factor at the $\Gamma$ and $K$ points of the Brillouin zone respectively. 
In contrast, in the CMF solution, one can directly evaluate the local order parameters given by the expectation values of the spin and quadrupolar operators. This is because in the CMF, the mean-field coupling acts as an external environtment that spontaneously breaks some of the symmetries of the ground state, similarly to what happens in the thermodynamic limit.

The results of Figure \ref{Fig:Figure3} show that some of the phase boundaries, in particular, the transitions to a Ferromagentic state and also the symmetry point $\theta=0.25\pi$ are very robust and do not depend on the method used to find the ground state. In contrast, other phase boundaries such as $\theta_{XXZ}$ and $\theta_{BB}$ have a strong dependency on the method and cluster size used. In the spin-$1$ model the results obtained by ED and CMF  are in good agreement when performing finite size scaling 
\cite{Moreno-Cardoner14, Lauchli06}. In contrast, in the spin-$1/2$ model CMF clearly beats ED and yields a value of $\theta_{XXZ}$ that approaches to the QMC result. The SS1 phase obtained by ED is indeed  restricted to a very tiny region in the phase diagram, due to finite size effects in this method.

\section{Entanglement characterization of triangular spin lattices}
Here we aim at characterizing the ground states of the different many-body phases of the previous models by studying  their entanglement content. We make use of several tools and concepts borrowed from quantum information theory. There are many quantities that can certify the presence of quantum correlations in a given system but not all provide the same information. Some of them act as a measure able not only to certify but also to quantify the amount of entanglement between two blocks of a system, others are not really measures of entanglement since they fail to not increase with respect to local operations performed on the spins. 

Notice that for the ground state, the information contained in the Schmidt decomposition is, by default, complete with respect to the corresponding bipartite splitting. A widely used quantity to compute entanglement for bipartite cuts is the entanglement entropy defined as $EE=-\sum_{i}\lambda_{i}\log\lambda_{i}$. Closely related is the so-called negativity \cite{Karol98}, which corresponds to the sum of the negative eigenvalues obtained after partially transposing with respect to one of the subsytems $\rho_{AB}^{T_{A}}$ of the given partition $A/B$. For pure states, the negative eigenvalues of the partial transposition are simply given the $-\lambda_{i}\lambda_{j}$ values. 

Yet another measure that can quantify entanglement, but which does not depend on the chosen cut, is the geometric entanglement. This can be defined as the maximal overlap that the exact  ground state $\ket{\psi}_{G}$ has with a separable state, i.e. $max_{PS}(\langle{\phi_{S}} \ket{\psi})$, where the maximization is performed over all possible (unentangled) product states (PS).

\section{Results}
\subsection{Entanglement Entropy}
\begin{figure}[H]
\includegraphics[width=0.65\textwidth,angle=0]{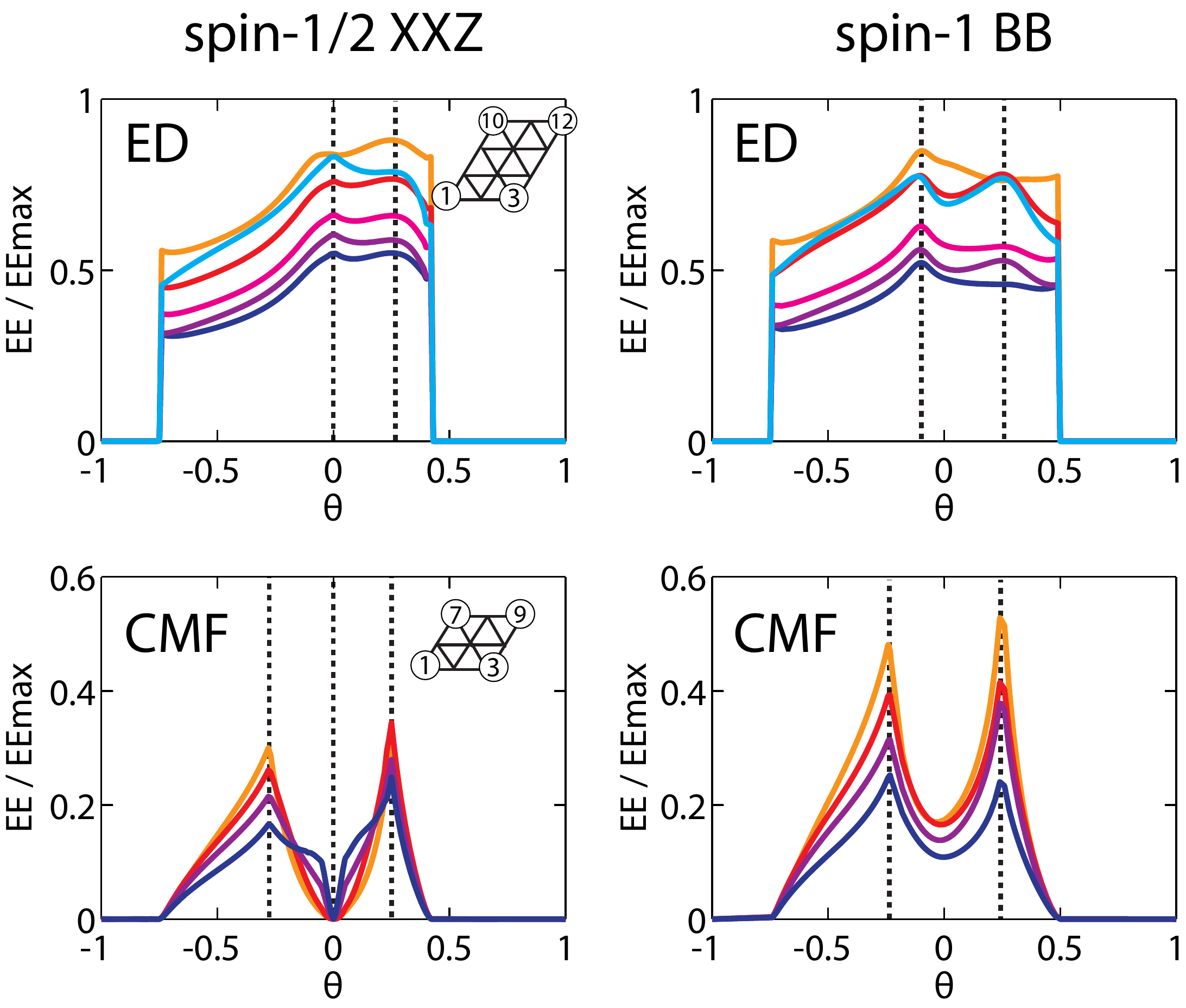}
\centering
\caption{(Color Online). Ground state entanglement entropy (EE) for different bipartitions renormalized to its maximum value (see text) for the spin-$1/2$ XXZ (left) and spin-$1$ BB (right) models. Top: results from ED for 12 sites. The partitions correspond to: 1-2-3 (orange)/ 1-4-7-10 (red)/ 1-2-3-4-5 (pink)/ 1-2-3-4-5-7 (purple)/ 1-2-3-4-5-6 (dark blue)/ 1-5-9-12 (cyan). Bottom: results from CMF of 9 sites. The partitions correspond to 1 (orange)/ 1-2 (red)/ 1-2-3 (purple)/ 1-2-3-4-5 (cyan). The dotted lines denote some of the phase boundaries. Note the different scale in the CMF results.}
\label{Fig:Figure4}
\end{figure}

We start by analyzing the entanglement entropy, that is, the Von-Neumann entropy given a partition of our system into two blocks $A,B$. This measure, although it can signal continuous quantum phase transitions as a logarithmic divergence on the size of the partition, does not reveal some information about the exact nature of the many-body phase, as it is for instance, the symmetry of the state or if it is locally ordered. Here we calculate and compare this quantity  for different bipartitions, both for the ground state obtained from exact diagonalization (ED), as well as for the ground state obtained from cluster mean field approach (CMF). Our results are shown in Figure \ref{Fig:Figure4}. 

For the exact diagonalization, given the relatively small system sizes (up to 12 sites), and because we are dealing with a two-dimensional system, the way the cut is done becomes crucial. In order to compare different partitions, we renormalize the EE to its maximum value for each bipartition, given by $EE_{\max}= \log(\dim)$, where $\rm{dim}$ refers to the Schmidt rank. In the cases we study this coincides with its maximum value $\dim=\min(d^{L_A},d^{L-L_A})$, being $L_A$ and $L$  the number of sites of the partition and of the whole system respectively.  As expected, even for small systems, the EE displays a local maximum at the points corresponding to the boundaries of continuous phase transitions. In contrast, for the first order transitions to FM phases, the EE is discontinuous. Notice also that in the studied phases, the relative value of entanglement content (with respect to the Schmidt rank associated to the partition) decreases in general with the Schmidt rank. However, some particular partitions (as for instance, the one containing all sites belonging to the same sublattice (cyan line in Figure \ref{Fig:Figure4})) show larger entanglement entropy in the spin-$1/2$ model. 
This seems natural in a locally ordered phase with a sublattice structure, where quantum correlations are maximal between sites with different sublattice order. 

It is interesting to see how entanglement is incorporated within the CMF approach. Although the quantitative behavior of EE is very different, it still shows some analogies with the ED results, as for instance the presence of local maxima at the continuous phase boundaries. The increase of this quantity when approaching the phase transition is however, much more abrupt than in the ED results. This has the only exception of the transition between the two supersolid phases (SS1 and SS2) in the spin-1/2 model, where it is minimum. Nevertheless, and interestingly enough, the EE increases at the two SS phases with the Schmidt rank of the partition. 
 
\subsection{Schmidt Decomposition}
\subsubsection{General features: Entanglement Spectrum (ES) and Schmidt Eigenvectors (SE)\\}
The Schmidt decomposition of a pure state is given by  $\psi_{GS}=\sum \sqrt{\lambda_{i}}\ket{u_{i}^{A}}\ket{v_{i}^{B}}$, being  $\{ \ket{ u_{i}^{A}},\ket{v_{i}^{B}}\}$ the Schmidt eigenvectors (SE) forming two biorthonormal basis. The entanglement spectrum (ES) is nothing else than their weights in the Schmidt decomposition. For a pure state and a particular partition, these quantities contain all the information of the ground state. Despite the information encoded in the SE might be difficult to extract, studying the behavior of their weight for different partitions can reveal further information.

Let us first analyze the results obtained by exact diagonalization (ED) after imposing boundary conditions as discussed in Section \ref{Sec:Models}. The choice of these boundary conditions crucially determines the symmetry properties of the ground state. Our results for a cluster of 12 sites are shown in Figure \ref{Fig:Figure5} and Figure \ref{Fig:Figure6} for spin-$1/2$ and spin-$1$ respectively. Latter on we analyze the same quantities for the many-body state obtained by using the CMF approach. These are shown in Figure \ref{Fig:Figure7}.

At first sight, the ES of a two-dimensional system shows a complicated dependence on the Hamiltonian parameters in the vicinity of the quantum phase transitions as compared to one-dimensional cases. However, significant features can be extracted: quite generally the Schmidt coefficients show many crossings, and the number of crossings and levels depend obviously on the chosen partition. We find that as long as the ground state is non-degenerated, the Schmidt rank is maximal, and the different Schmidt eigenstates continuously vary and swap at each of the crossings. This is true not only inside a phase, but also at the points associated to a continuous second order phase transition in the thermodynamic limit, as for instance in the SS2-AFQ transition point ($\theta=0.25\pi$) in the spin-$1/2$ XXZ model, and in the FQ-AFM ($\theta=-0.11 \pi$)  and AFM-AFQ ($\theta=0.25 \pi$) transitions in the spin-$1$ model (see Figure \ref{Fig:Figure1}).  Moreover, the crossings between the dominant SE take place close to the phase boundaries, and we expect that as the system size increases they will approach to the corresponding critical points. Between different phases, we observe quite generally that the dominant Schmidt eigenvectors display already the ordering characteristic of the phase. This can be asserted by looking at the corresponding structure factors obtained from the Schmidt eigenvectors. Thus, it is not surprising that they cross near a phase transition. It would be very convenient to perform a similar analysis for larger cluster sizes.

In contrast, the eigenstates are in general discontinuous when crossing an energy degeneracy point, as it happens in the spin-$1/2$ XXZ model at the Ising point ($\theta=0$), as shown in Figure \ref{Fig:Figure5}. In this case, the two phases cannot be adiabatically connected (at least for the sizes studied), and not only the entanglement spectrum but also the Schmidt eigenvectors present a discontinuity. This also holds in first order phase transitions when entering the ferromagnetic (FM) phases.

In the CMF solution the associated entanglement spectrum is very different. There is mainly a dominant Schmidt eigenstate that possess the same local ordering that characterizes the phase. The ES and SE are continuous, but not the corresponding derivatives, clearly signaling the different phase boundaries.\\

\subsubsection{Analysis of the Degeneracy in the ES and characterization of the SE\\} 
\begin{figure}[H]
\centering
\includegraphics[width=0.65\textwidth,angle=0]{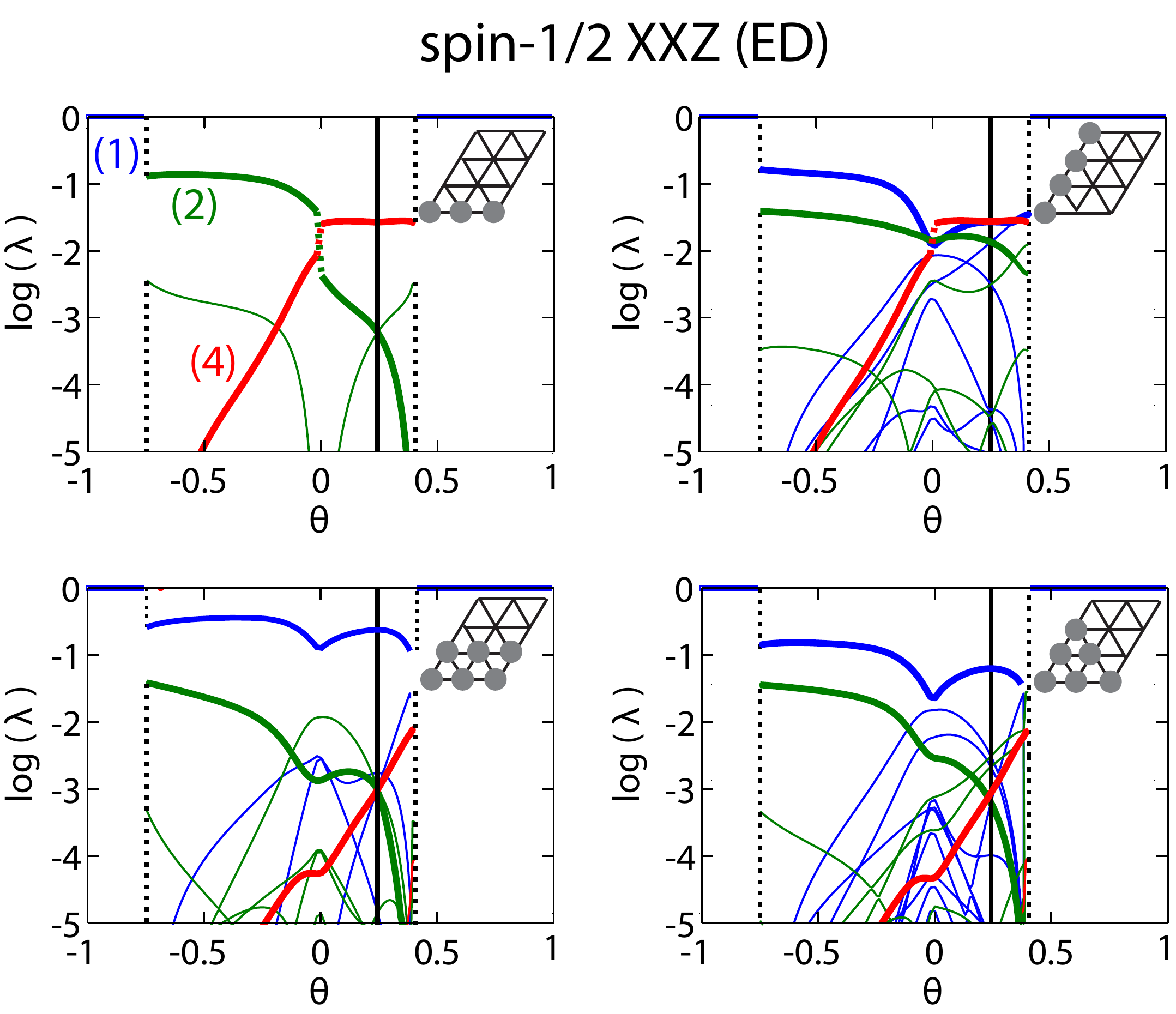}
\caption{Entanglement spectrum for the spin-$1/2$ XXZ model obtained by ED for different partitions as denoted in the inset of each figure. The number of levels in the ES is given by the subsystem size. Different colors correspond to different degeneracies as numbered in the plot. The different levels correspond to Schmidt eigenstates (SE) with well defined $S^z$. Non-degenerated levels correspond to SE with $S^z=0$, higher degeneracies correspond to SE with $S^z=\pm M$, $M\neq 0$. At the SU(2) point (black line) the number of crossings is larger and the SE have well defined $S^2$ (see text).} 
\label{Fig:Figure5}
\end{figure}

\begin{figure}[H]
\centering
\includegraphics[width=0.65\textwidth,angle=0]{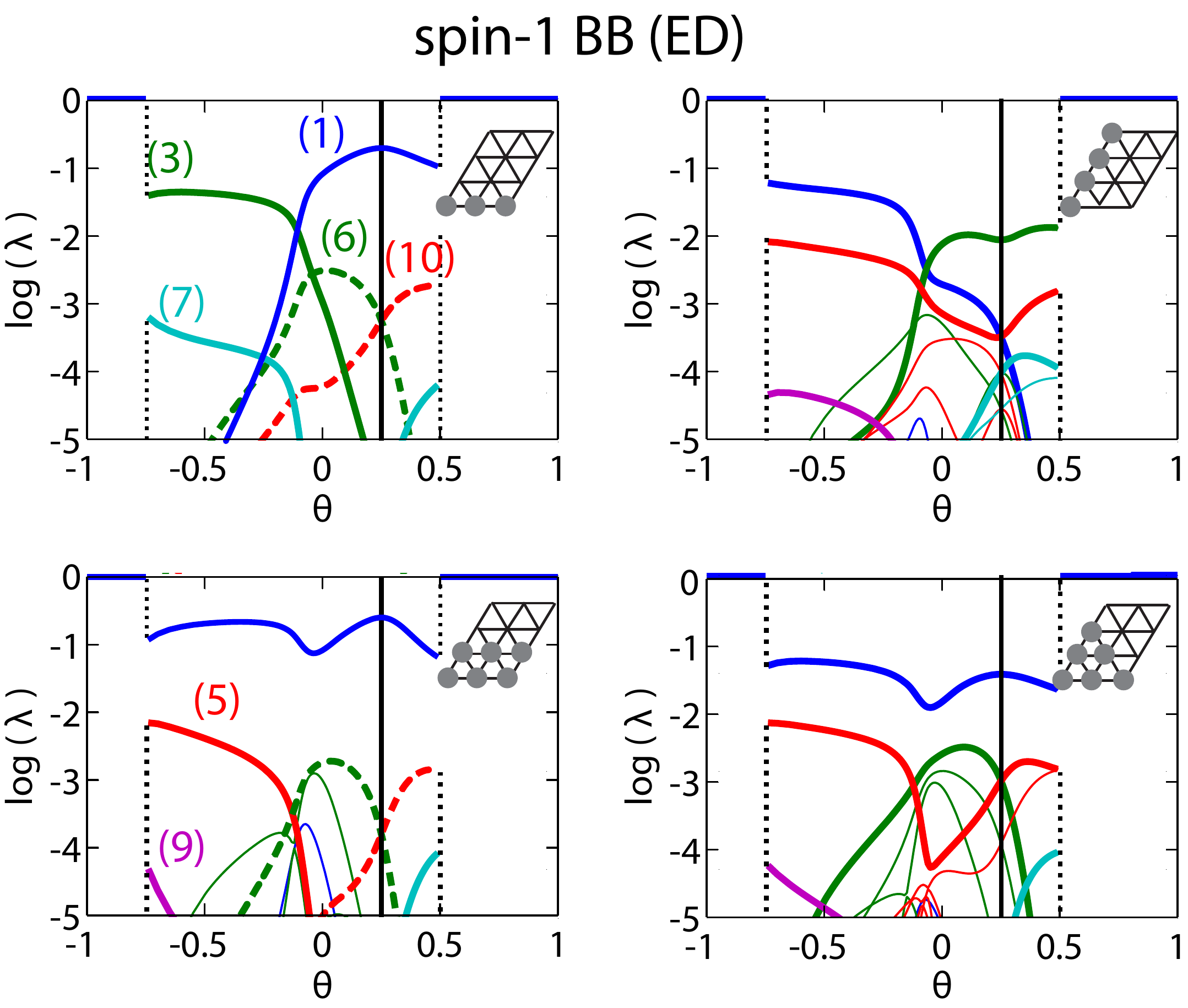}
\caption{Entanglement spectrum for the spin-$1$ BB model obtained by ED for different partitions as denoted in the inset of each figure. The number of levels in the ES is given by the subsystem size. Different colors correspond to different degeneracies as numbered in the plot. Solid (dashed) lines correspond to Schmid eigenvalues with angular momentum $j$ ($j$) and degeneracy $\nu_j=(2j+1)$ ($\nu_j=2(2j+1)$) (see text).  At the SU(3) point (black line) the number of crossings is larger (see text).} 
\label{Fig:Figure6}
\end{figure}

\begin{figure}[H]
\centering
\includegraphics[width=0.65\textwidth,angle=0]{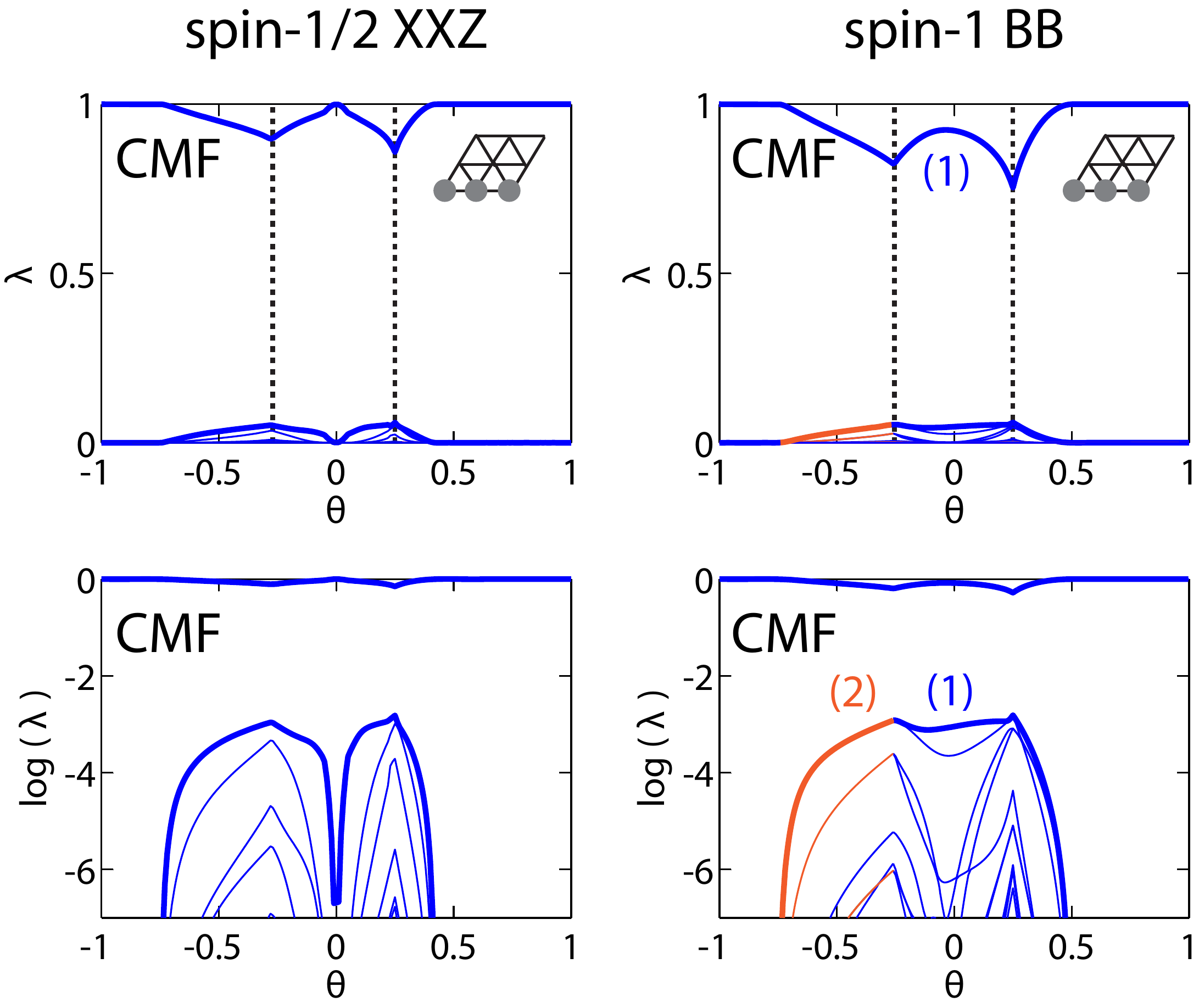}
\caption{Entanglement spectrum for the spin-$1/2$ XXZ (left) and spin-$1$ BB (right) models obtained by CMF for a particular partition as denoted in the inset of the figures.  Different colors correspond to different degeneracies as numbered in the plot. Top (linear scale): the first level dominates and the minimum signals continuous phase transitions. Bottom (log scale): the subdominating levels are singly-degenerated in most phases except in the FQ phase (see text).} 
\label{Fig:Figure7}
\end{figure}

For all the analyzed phases of these models, the different levels are grouped into multiplets with a degeneracy that does not change across the boundaries associated to continuous phase transitions. This is in stark contrast to the Dimerized-Haldane phase transition in a one-dimensional chain, where the degeneracy changes when approaching the transition \cite{XGWen11,Lepori13}. Here the degeneracy of these multiplets is directly linked to the local symmetry of the ground state and to the bipartition considered. The different degeneracies are indicated in Figures \ref{Fig:Figure5} and \ref{Fig:Figure6} by using different colors. 

Indeed, suppose that the ground state possess a local symmetry that can be associated to a local unitary operator $U$ (that is, the ground state only acquires a phase under a transformation $U_G=U^{\otimes L}$, where $L$ is the number of sites). Then, it is clear that not only the density matrix will be invariant under this transformation, but also any reduced density matrix, thanks to the local character of the transformation. Note however, that when the ES is degenerated, the Schmidt eigenstates are not necessarily invariant under the transformation and can break that particular symmetry.\\ 

For instance, let us consider the spin-$1/2$ XXZ Hamiltonian, which possess $SO(2) \times Z_2$ symmetry, and so it does the ground state when it is non-degenerated. That is, for a $L$ sites cluster, the ground state is an eigenstate of the time reversal operator $T_L=\prod_{l\in L}  \sigma^y_l$ and also of the $\hat{z}$-component total angular momentum operator $S^z_{L}=\sum_{l\in L} S^z_l$, except in  the FerroMagnetic phases that are degenerated and can break these symmetries. Moreover, at $\theta=\pi/4$ the Hamiltonian is the Heisenberg model and the ground state has an additional $SU(2)$ symmetry. These symmetries are reflected in the Schmidt decomposition in the following way: the different levels of the ES correspond to different eigenstates of $S^z_{L_A}$ with eigenvalue $j_z$, where now $L_A$ is the subsystem size after the partition. When $j_z \neq 0$, then these levels are necessarily at least doubly degenerated, to ensure that the reduced density matrix has indeed $Z_2$ symmetry. In this case, there are two Schmidt eigenstates associated to each level, one with $j_z = m$ and  the other one with $j_z = -m$. Finally, at the $SU(2)$ point, the degeneracy of the levels is larger and the SE will have a higher symmetry, as discussed below.\\

We turn now into the spin-$1$ Bilinear-Biquadratic Hamiltonian, which is rotationally invariant. The ground state is now $SU(2)$ invariant, except in the Ferromagnetic phases. This is equivalent to say that the Hamiltonian and density matrices commute with $(\vec{S}_L )^2$, where $\vec{S}_{L}=\sum_{l\in L} \vec{S}_l$ is the total angular momentum operator. Thus, the Schmidt eigenstates have now well defined angular momentum $j$, and they correspond to integer irreducible representations of $SU(2)$, with multiplicity $\nu_j=2j+1$. Given a subsystem with $L_A$ sites, the maximum spin contributing is clearly $j_{max}=L_A$. In general, states with well defined angular momentum are not necessarily eigenstates of the permutation operator between two sites. However, we expect that the non-degenerate ground-state is also an eigenstate of the translation operator along the axis where periodic boundary conditions are imposed, and that the density operator is invariant under this transformation. If the cut preserves translational symmetry along this axis (as in bipartitions 1-2-3 and 1-2-3-4-5-6, see Figure \ref{Fig:Figure6}), this is then reflected into an invariant reduced density matrix. In this case some of the levels associated to eigenstates with angular momentum $j$ show larger degeneracy $\nu_j=2(2j+1)$. They correspond precisely to states that do not possess the symmetry, and need to be combined to yield an invariant density matrix. 

Finally, the $SU(3)$ symmetric point $\theta=\pi/4$ will show for any partition many crossings. At this point the Schmidt eigenstates are representations of the $SU(3)$ group with degeneracy given by the two corresponding Casimir operators $p$ and $q$, $\nu_{p,q}=(p+1)(q+1)(p+q+2)/2$.\\ 

Finally, we have analyzed the multiplicities of the levels of the ES in the CMF solution. An important effect of the mean-field coupling with other clusters (which might act as an environment) is that some symmetries of the state will be spontaneously broken, similarly to what happens in the thermodynamic limit. This is clearly reflected in the ES, where most of the Schmidt eigenstates are non-degenerated. However, not all the symmetries will be totally broken. For instance, a FQ phase will still show a rotational symmetry around the axis defined by the vector director. Indeed, we have found that for this phase some of the levels in the ES are doubly degenerated, and that they precisely coincide with the Schmidt eigenstates that do not preserve this symmetry. These eigenstates have anisotropic fluctuations in the normal plane to the vector director, and thus, double degeneracies are needed to restore $SO(2)$ rotational invariance in the reduced density matrix.\\

With all these results, we thus can conclude that the multiplicities of the ES reveal information about the local symmetries of the ground state.\\

\subsection{Geometric Entanglement}
All previous measures can detect entanglement when present, but its exact value depends on the bipartition that has been  chosen. Here we study a different measure that can quantify entanglement in a more general way, which does not depend on the details of the chosen cut. The so-called geometric entanglement can be defined as the minimal distance (or maximal overlap) between the strongly correlated state $\ket{\psi}$ and its closest separable approximation $\ket{\phi_S}$, that is $GE=1-\max_{PS}(\langle{\phi_{S}} \ket{\psi})$ where the maximization is performed over all possible product states (PS). 

\begin{figure}[H]
\centering
\includegraphics[width=0.65\textwidth,angle=0]{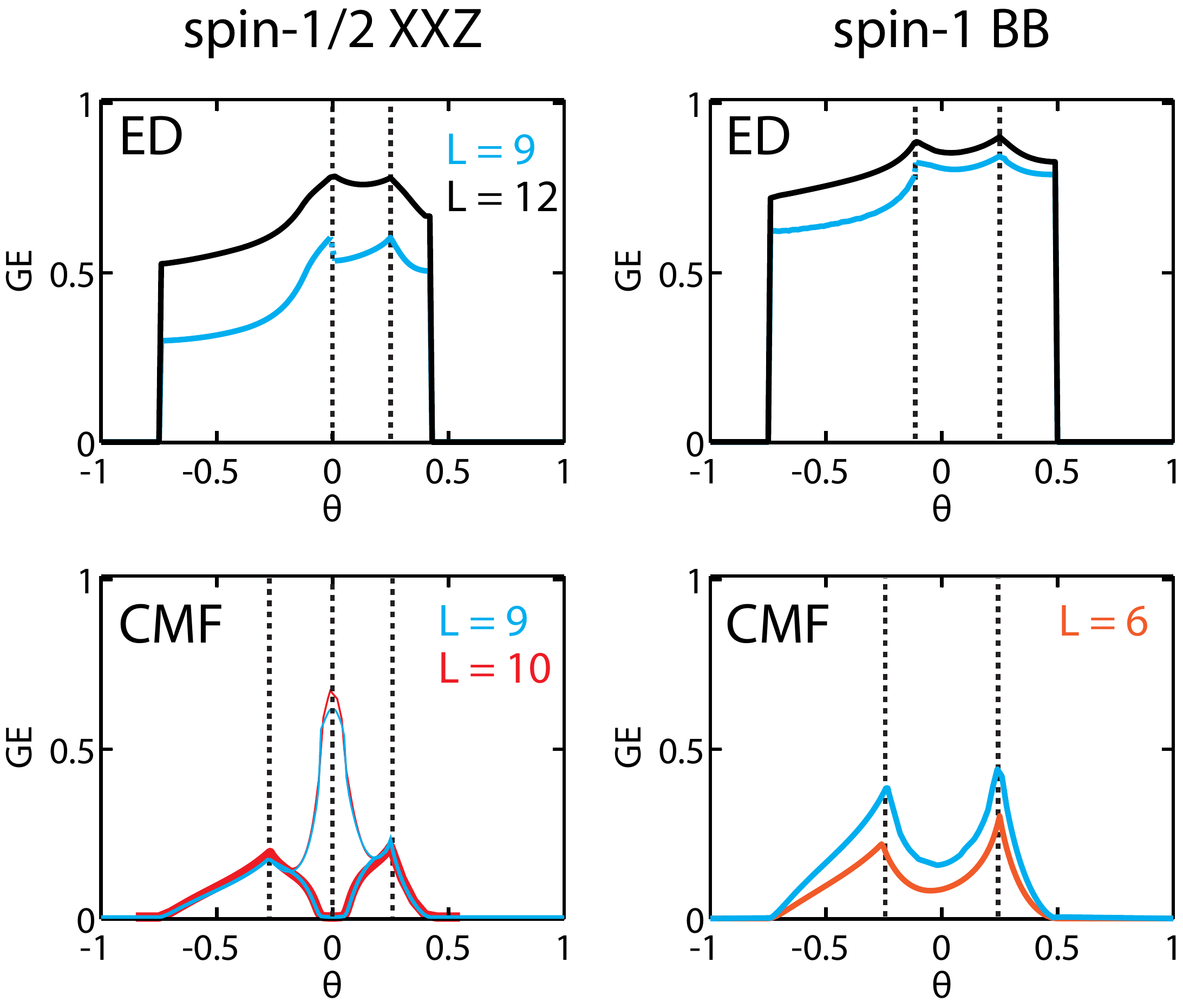}
\caption{Geometric entanglement (GE) for different cluster sizes for the ground state of the two models: spin-$1/2$ XXZ (left) and spin-$1$ BB (right). Top: results from ED for 9 and 12  sites. Bottom: results from CMF for 6, 9 and 10 sites. The GE shows maxima at the phase boundaries and a discontinuity in its derivative. In the CMF solution for the spin-$1/2$ model only  the solution showing three-sublattice order has vanishing entanglement near the Ising point (thick line), whereas for the other solutions the GE abruptly increases (thin line) (see text).}
\label{Fig:Figure8}
\end{figure}

We numerically have found this quantity for the ground state obtained by ED and CMF for the two models, and compare the results between different cluster sizes. They are shown in Figure \ref{Fig:Figure8}. The GE presents a similar qualitative behavior as the EE, in particular, it also presents a local maxima at the phase boundaries of continuous phase transitions. However, these peaks are sharper than those in the EE, and the derivative with respect to the parameter $\theta$ is discontinuous, in contrast to the EE results. 

For all the studied phases, the closest separable state presents always the three-sublattice structure. Moreover, in the CMF approach this state coincides with the Gutzwiller ansatz, that is, with the separable state that minimizes the mean energy. In the Supersolid phases in the XXZ model, where not all cluster solutions possess this sublattice pattern, this is only true for the solution with three-sublattice structure. For the other two solutions the overlap becomes smaller when approaching the Ising point ($\theta=0$), as shown in Figure \ref{Fig:Figure8}. 

\section{Conclusions}
In summary, we have analyzed the entanglement content of the ground state of triangular spin lattice models both near quantum phase transitions and within a given phase. To this aim we have used exact diagonalization techniques as well as the cluster mean field approach.

On the one hand, for exact diagonalization results, we observe that if the ground state is non-degenerate, the symmetries of this state are faithfully reflected into the Schmidt levels, and their footprints are degeneracies. In this case, where ground states are adiabatically connected from one phase to another, the Schmidt levels are all continuous and quantum phase transitions are signaled by {\it {crossings}} between the dominant Schmidt levels. Due to finite size effects, these crossings of the Schmidt eigenvalues do not necessarily coincide with critical points, and therefore the results obtained from the entanglement spectrum (and all derived quantities such as Renyi entropies) require finite size scaling methods to converge to the exact value where the quantum phase transition occurs. For first order phase transitions the degeneracy of the ground state leads immediately to a discontinuity on the Schmidt eigenvalues, and there is not an adiabatic connection between ground states. Finally, we observe that phase boundaries associated to an extra symmetry of the Hamiltonian (like SU(2) in spin-$1/2$ or SU(3) in spin-$1$) are well indicated by a massive crossing of many Schmidt levels at that point, independently from the size of the system. 
On the other hand, the analyzed entanglement measures are quantitatively very different in the cluster mean field solution, but they still show some similarities with the exact diagonalization results.  The mean field coupling between clusters acts effectively as noise for the quantum bonds within the cluster, and tends to smooth all entanglement spectrum features. In this case, symmetries are often broken as it happens in the thermodynamic limit, and the reduced density matrix is not forced to keep all possible terms to preserve those symmetries. Within this approach, the Schmidt gap (or difference between the two largest Schmidt eigenvalues) tends to close when approaching a phase transition \cite{DeChiara12}, but the exact location of the phase boundary is again affected by the finite size of the clusters. 

Finally we also compare global entanglement measures, such as entanglement entropy and geometric entanglement, both for the exact diagonalization and the cluster mean field  solutions. In the latter, and as expected, phase transitions are much strongly signaled than in the exact diagonalization result. Geometric entanglement is an entanglement measure not based on any partition of the system, but which it rather computes how far the ground state of a strongly correlated system is from a product state. This measure detects very well phase transitions, and interestingly enough, it yields the Gutzwiller mean field ansatz as the closest product state to the cluster mean field approach. 

To conclude, our study shows that in 2D systems also entanglement measures are very sensitive to critical behavior, and we demonstrate that the degeneracies of the Schmidt eigenvalues reflect the symmetries of the ground state. 

 \ack
We acknowledge useful discussions with E. Mascarenhas and J.M. Escartin.  Financial support from: MINECO FIS2008-01236 (Spain), Grant SGR2009-1289 (Catalonia),  Grant 618074 (EU Project TherMiQ), 
Grant 43467 (Templeton Foundation), Grant EP/L005026/1 (EPSRC), 
Grant EP/K029371/1, S.P. acknowledges partial support from MCTI and UFRN/MEC
(Brazil).

\end{document}